\documentclass{cpbtex}

\usepackage[dvipsnames]{xcolor}
\usepackage[normalem]{ulem} % use for strikethrough
\usepackage{multirow} % for table
\usepackage{graphicx}
\usepackage{tabularx}
\usepackage{caption}
\usepackage[main=english]{babel}
\usepackage[warnundef]{jabbrv}

\usepackage{hyperref}
\hypersetup{
    colorlinks=true,  % 开启链接颜色（代替方框）
    linkcolor=black,  % 将内部链接（如目录、公式）设为黑色
    citecolor=black,  % 将参考文献链接设为黑色
    urlcolor=black    % 将网页链接设为黑色
}

\addto\captionsenglish{}

\begin{document}

\title{\Large Topological property of graphene with  triangular array of nanoholes}

% Title should be concise; avoid abbreviations if possible; and not begin with `A', `An', `The', or `Study on'.

\author{Yong-Cheng Jiang(江咏城)$^{1,2}$, \ Xing-Xiang Wang(王星翔)$^{1}$, \ and \ Xiao Hu(胡晓)$^{1,2}$\thanks{Corresponding author. E-mail:~hu\_xiao@shu.edu.cn}\\
$^{1}${Institute for Quantum Science and Technology, Shanghai University,}\\
{Shanghai 200444, China} \\
$^{2}${Department of Physics,  College of Sciences, Shanghai University,}\\ 
{Shanghai 200444, China}
}  % The line break was forced via \\

\date{\today}

\maketitle

\begin{abstract}

The nontrivial band topology for graphene with regular arrays of nanoholes with $C_{6v}$ symmetry is investigated theoretically. For the case of $3\sqrt{3}\times 3\sqrt{3}$ triangular array of nanoholes, we find an energy gap at $\Gamma$ point around the Fermi level associated with a band inversion which induces change in parity indices, whereas deep below the Fermi level there are a bunch of valence bands (VBs) characterized as obstructed atomic limit (OAL) which also accommodate imbalance in parity indices. This band structure renders the gap at the Fermi level topologically trivial and  carrying no edge states, while the nontrivial band topology of the OAL manifests in two flat bands in the ribbon structure associated with localized electronic states at ribbon edges. The present results exhibit rich topological behaviors in graphene derivatives waiting for explorations.
  
\end{abstract}



\section{Introduction}

Honeycomb lattice plays a special role in fostering topology physics~\cite{Klitzing1980}. Haldane model~\cite{Haldane1988} was defined on honeycomb lattice where the nearest-neighboring (nn) hopping integrals form the Dirac dispersion, and the complex next-nearest-neighboring (nnn) ones open a gap establishing the Chern insulator~\cite{Thouless1982,Weng2015}. Kane and Mele noticed that, considering the spin degree of freedom, spin-orbit coupling appears associated with the nnn hopping integral rendering the topological insulator characterized by $Z_2$ invariant~\cite{Kane2005Z2,Kane2005Gr}.

A spinless tight-binding (TB) model was proposed on honeycomb lattice~\cite{Wu2015,Wu2016} where real nn hopping integrals are detuned in a way respecting $C_{6v}$ symmetry, known as Wu-Hu model~\cite{Leykam2026,Zhu2026}. When the hopping integral inside a hexagonal unit cell $t_0$ is smaller than the one between unit cells $t_1$, namely $t_1>t_0$, a topological energy gap $\delta=t_1-t_0$ is opened in the Dirac dispersions at $\Gamma$ point of the Brillouin zone (BZ). The eigen wavefunctions at the upper/lower band edge are $p_x$ \& $p_y$/$d_{x^2-y^2}$ \& $d_{2xy}$. This $p-d$ band inversion induces an imbalance in the parity indices of VBs (VBs) between the high-symmetry momenta $\Gamma$ and M points, yielding nontrivial topology fostered by crystalline symmetry\cite{Fu2011}, which can be characterized in terms of Euler class when $C_2 T$ symmetry is exploited over the entire BZ~\cite{Palmer2021,Wang2025,Ahn2019,Wang2020}. 

The above physics can be extended to graphene with regular arrays of nanoholes~\cite{Kariyado2018,Zhao2024,Shima1993,Pedersen2008,Bai2010}, which may be fabricated using bottom-up synthesis methods~\cite{Moreno2018,Song2025}. Generally when $m\sqrt{3}\times m\sqrt{3}$ and $3m\times3m$ triangular and/or honeycomb array of nanoholes are punctured in graphene, where a nanohole is formed by removing six carbon atoms with appropriate hydrogenation, an energy gap is opened in double Dirac cones at the Fermi level by the intervalley interaction due to the BZ folding. 

The topology of a gapped state can be characterized by localization at non-atomic Wyckoff positions of Wannier functions associated with relevant bands~\cite{Bradlyn2017,Cano2018,Marzari2012}. As a matter of fact, topological materials have been diagnosed recently with the aids of  material databases~\cite{Zhang2019,Vergniory2019,Tang2019,Tang2019NP} using topological quantum chemistry~\cite{Bradlyn2017,Cano2018} and symmetry indicators~\cite{Po2017}. 

In this work we focus on graphene with $3\sqrt{3}\times 3\sqrt{3}$ triangular array of nanoholes. For this system a topology diagnose can be performed based on analysis on Wyckoff positions of Wannier centers directly from the number of VBs in modulo six related to the $C_{6v}$ symmetry, $N_{\text{VB}} \pmod{6}$ \cite{Jiang2026JPSJ}. Interestingly, we find topological properties in this system although the diagnose in terms of $N_{\text{VB}} \pmod{6}$ suggests a trivial topology at the Fermi level.

\section{Tight-binding model}

 	\begin{figure}[t]
		\centering
		\includegraphics[width=0.75\textwidth]{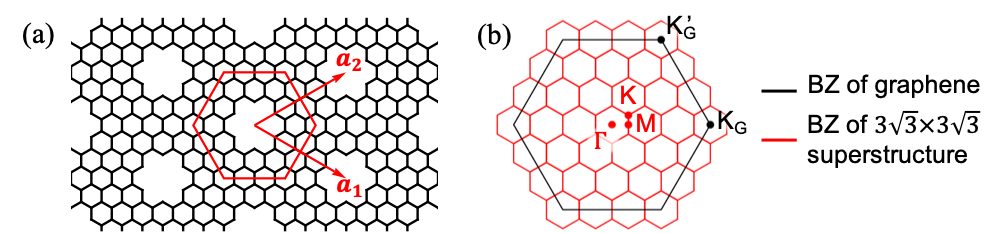}
		\caption{(a) Schematic structure for graphene with $3\sqrt{3}\times3\sqrt{3}$ triangular array of nanoholes with its unit cell and unit vectors.
		(b) Brillouin zone (BZ) folding from pristine graphene to graphene with $3\sqrt{3}\times3\sqrt{3}$ superstructure.
		}
		\label{fig:structure}
	\end{figure}

    As illustrated in Fig.~\ref{fig:structure}(a), the system studied in this work is formed by introducing a $3\sqrt{3}\times3\sqrt{3}$ triangular array of nanoholes in graphene, where each nanohole is formed by removing six removing carbon atoms with perimeter passivated by hydrogen. We denote prinstine graphene as the $1\times1$ structure for its two unit vectors with the lattice constant $a_\text{Gr}$, so that $3\sqrt{3}\times3\sqrt{3}$ superstructure represents that the lattice constants of two unit vectors are $|\bm{a}_{1,2}|/a_\text{Gr}=3\sqrt{3}$.  
    We adopt the following TB Hamiltonian to describe the electronic properties of of $\pi$ electrons for the superstructured graphene:
    \begin{equation}\label{eq:Hamiltonian}
        H = - t \sum_{\langle i,j \rangle} c^{\dagger}_i c_j + \text{H.c.},
    \end{equation}
    where the sites in nanoholes are omitted in summation and the nn hopping energy $t=2.7$~eV~\cite{CastroNeto2007} is assigned for the remaining sites.

\section{Results}

\subsection{ Nontrivial topology}

 	\begin{figure}[t]
		\centering
		\includegraphics[width=0.75\textwidth]{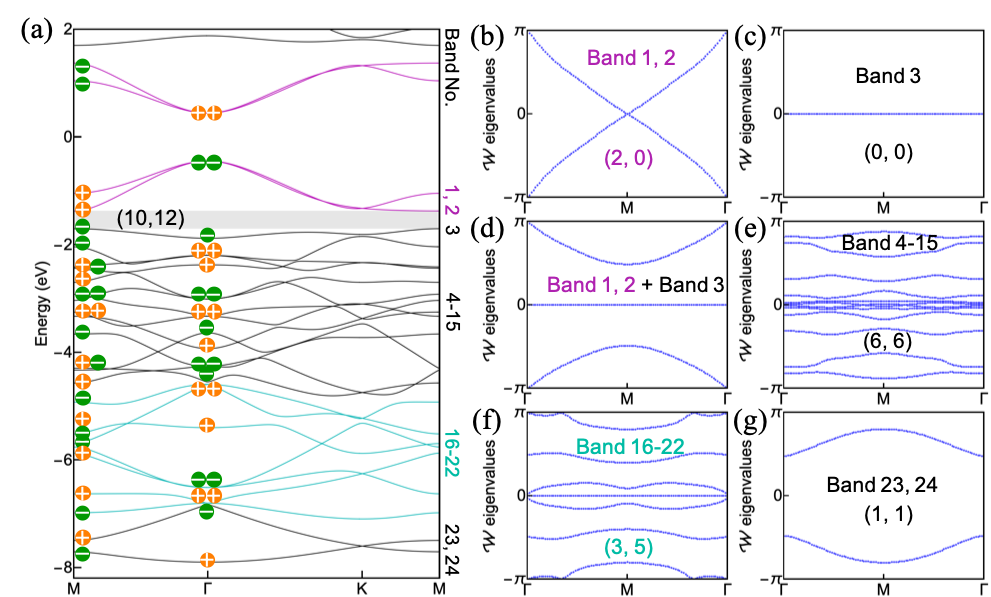}
		\caption{(a) Band structure with the top two valence bands and bands No.~16--22 characterized by fragile topology and obstructed atomic limit, respectively.
		Parity indices at the M and $\Gamma$ points accumulated below the energy gap in gray are denoted in the bracket.
		(b)--(g) Eigenvalues of Wilson-loop phases for bands No.~1,~2 (b), No.~3 (c), No.~1--3 (d), No.~4--15 (e), No.~16--22 (f) and No.~23,~24 (g), parity indices within each set of bands at the M and $\Gamma$ points are denoted in the bracket.
		}
		\label{fig:bulk}
	\end{figure}

    The band structure for graphene with $3\sqrt{3}\times3\sqrt{3}$ triangular array of nanoholes calculated by the TB model~(\ref{eq:Hamiltonian}) is shown in Fig.~\ref{fig:bulk}(a), where all VBs with parity index~\cite{Kariyado2018} at the $\Gamma$ and M points are displayed explicitly. 
    Near the Fermi level, a large energy gap is opened in the double Dirac cones of pristine graphene, which is induced by intervalley interactions from the BZ folding depicted in Fig.~\ref{fig:structure}(b).
    The parity indices accumulated for all VBs between $\Gamma$ and M points are balanced, namely $(N_{\text{M}}^{+},N_{\Gamma}^{+})=(12,12)$, indicating trivial topology for the energy gap at the Fermi level.
    The trivial topology is expected from 
    $6d$ or $6e$ Wyckoff positions of Wannier centers since the number of VBs 
    $N_{\text{VB}}= (3\sqrt{3})^2-3$ modulo six is zero \cite{Jiang2026JPSJ}.

    Interestingly, by further characterizing the topology within the entangled bands below the Fermi level using Wilson loop~\cite{Yu2011}, we find two sets of bands carrying nontrivial topology, rather than all VBs being topologically trivial. 
	First of all, the eigenvalues of Wilson-loop phases for the top two VBs separated from other bands wind from $-\pi$ to $\pi$. It is however suppressed when the trivial band No.~3 is added as shown in Figs.~\ref{fig:bulk}(b)--(d), indicating fragile topology~\cite{Cano2018,Po2018}.
	Bands No.~16 through No.~22 are characterized as obstructed atomic limit (OAL)~\cite{Bradlyn2017,DePaz2019} as judged from unbalanced parity indices $(N_{\text{M}}^{+},N_{\Gamma}^{+})=(3,5)$ while the eigenvalues of Wilson-loop phases without winding (see Figs.~\ref{fig:bulk}(a) and \ref{fig:bulk}(f)).
	For other VBs, the eigenvalues of Wilson-loop phases display no winding and the parity indices are balanced (see Fig.~\ref{fig:bulk}(e) and \ref{fig:bulk}(g)), meaning that these bands are topologically trivial. 
    We notice that graphene with $3\sqrt{3}\times3\sqrt{3}$ triangular array of nanoholes is a case where VBs exhibit both fragile topology and OAL, and both types of topology may be characterized by a nontrivial $Z_2$ invariant such that their combination leads to a trivial energy gap at the Fermi level.

\begin{figure}[t]
		\centering
		\includegraphics[width=0.75\textwidth]{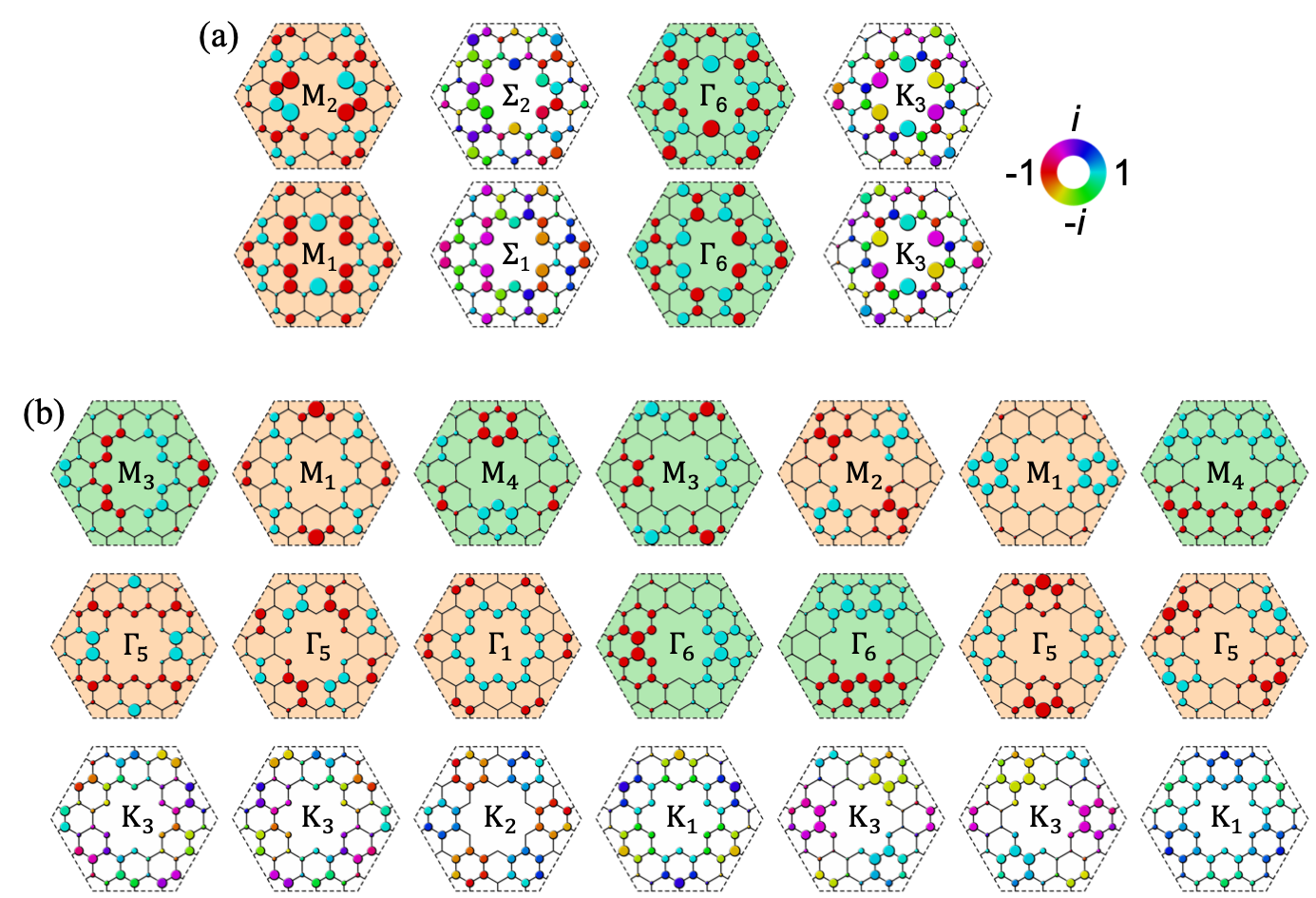}
		\caption{(a)~Wave functions of the top two valence bands at the high-symmetric momenta M, $\Sigma$, $\Gamma$ and K. 
	 	(b) Wave functions of bands No.~16--22 at the high-symmetric momenta M, $\Gamma$ and K displayed in the descending order of energy from left to right. 
	 	The amplitudes and phases of wave functions are denoted by size and color of dots, respectively.
		}
		\label{fig:WFs}
	\end{figure}

	The fragile topology and OAL can also be diagnosed from the irreducible representations (irreps)  at high-symmetry momenta~\cite{Bradlyn2017}. 
    In a $C_{6v}$-symmetric system, the irreps at high-symmetric momenta are summarized in Table~\ref{fig:table} in Appendix.
	For the top two VBs, the irreps for the wave functions are $(\text{M}_1,\text{M}_2,\Sigma_1,\Sigma_2,\Gamma_6,\text{K}_3)$ as displayed in Fig.~\ref{fig:WFs}(a), which characterize fragile topology by matching the difference of two elementary band representations, namely $(B_1)_{3c}\ominus(B_2)_{1a}$~\cite{Cano2018}.
    This band representation indicates an obstruction to construct exponentially localized  Wannier functions while respecting all symmetries of the system. However, after adding a trivial band associated with a localized Wannier orbital $B_2$ (see its symmetry in the top panel of Table~\ref{fig:table} in Appendix) at the $1a$ Wyckoff position, the resulting three bands can be represented by three localized Wannier orbitals $B_1$ at $3c$ Wyckoff positions.
	For bands No.~16 through No.~22, the irreps for the wave functions at the M, $\Gamma$ and K points are $(2\text{M}_1,\text{M}_2,2\text{M}_3,2\text{M}_4,\Gamma_1,2\Gamma_5,\Gamma_6,2\text{K}_1,\text{K}_2,2\text{K}_3)$ as shown in Fig.~\ref{fig:WFs}(b), which match the sum of two elementary band representations, namely $(E)_{2b}\oplus(A_1)_{3c}$~\cite{Bradlyn2017}. 
	This band representation has localized, symmetric Wannier functions sitting away from atomic sites, and thus its topology is characterized as OAL~\cite{Bradlyn2017}.    

\subsection{ Topological edge states}

\begin{figure}[t]
		\centering
		\includegraphics[width=0.75\textwidth]{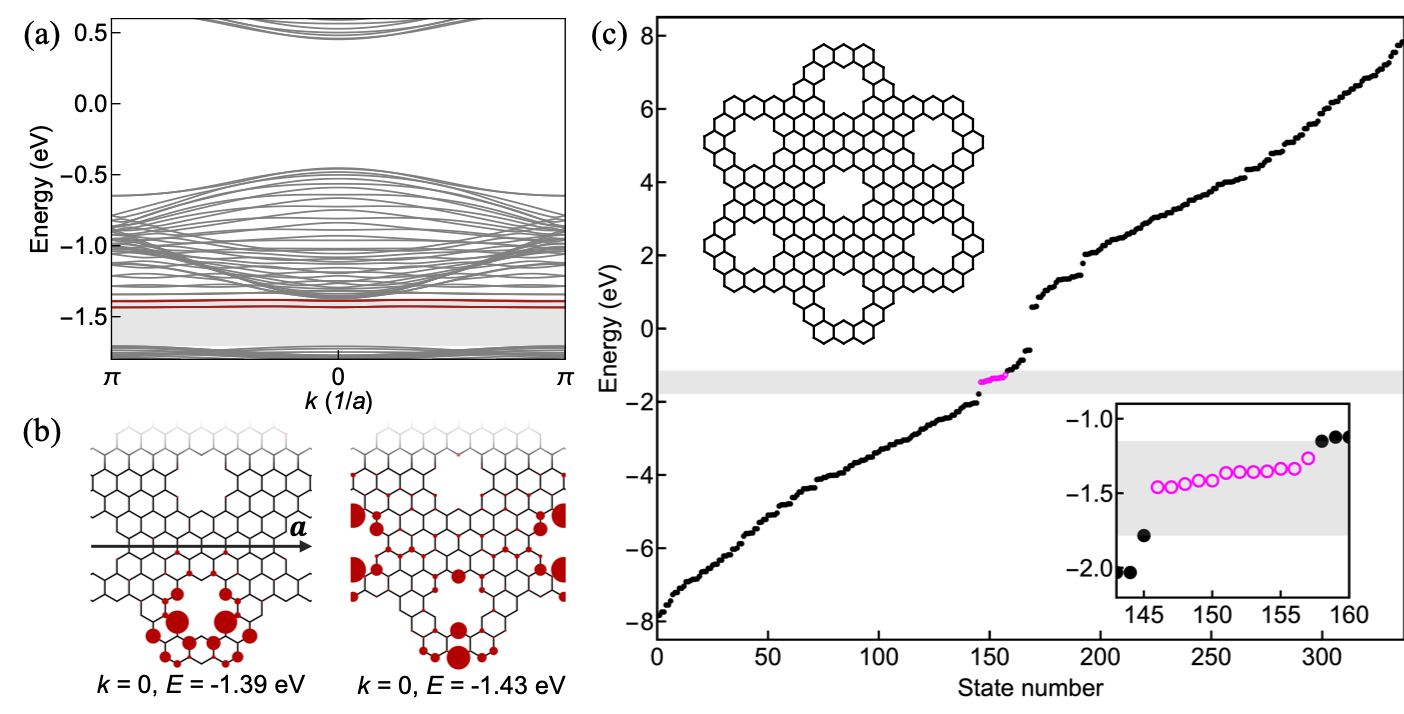}
		\caption{(a) Energy dispersion of a ribbon structure with the topological band gap denoted by a gray region.
		(b) Local density of states for the immobile edge states. 
		(c) Energy spectrum of a flake composed of seven unit cells, with edge states (magenta dots) in the energy gap (gray region) around $E=-1.5$~eV.
		The left inset is the schematic of the flake structure, and the right inset shows an enlarged view of the energy spectrum
		}
		\label{fig:TES}
	\end{figure}

    We display the energy dispersion for a ribbon structure which is infinitely long in $\bm{a}=\bm{a}_1+\bm{a}_2$ direction and of 20 unit cells along $\bm{a}_2$ direction in Fig.~\ref{fig:TES}(a). 
    Note that the edge morphology of the ribbon is \textit{molecular-zigzag}~\cite{Kariyado2017}, as shown in Fig.~\ref{fig:TES}(b), where hexagonal unit cells remain intact.
    Two immobile edge states, with their distributions disconnected in the infinite direction as illustrated in Fig.~\ref{fig:TES}(b), appear within a band gap of 0.3~eV near $E=-1.5$~eV (gray region in Fig.~\ref{fig:TES}(a)).
    The parity indices accumulated for the VBs below this band gap are unbalanced $(N_{\text{M}}^{+},N_{\Gamma}^{+})=(10,12)$, as shown in Fig.~\ref{fig:bulk}(a).
	We can see that the bands of OAL centered at
	$E=-6$~eV account for the imbalance of parity indices and thus the appearance of edge states in the band gap at $E=-1.5$~eV. 
	In contrast, although the top two VBs are charaterized as fragile topology, there is edge state at the Fermi level given that the parity indices accumulated by all VBs are balanced 
	$(N_{\text{M}}^{+},N_{\Gamma}^{+})=(12,12)$.  
	It is clear that the parity indices dominate the energy where the edge states appear.

	In the flake structure consisting of seven unit cells and respecting $C_{6v}$ symmetry as depicted in the left inset of Fig.~\ref{fig:TES}(c),
    we also find that the edge states associated with the bands of OAL appear in the energy gap at $E=-1.5$~eV  and no edge state exists at the Fermi level, as presented in Fig.~\ref{fig:TES}(c). 
    Note that this flake structure is the smallest one which can host edge states, providing a feasible experimental platform to check the edge states associated with fragile topology and OAL.

\subsection{ Design for molecular graphene}

\begin{figure}[t]
		\centering
		\includegraphics[width=0.6\textwidth]{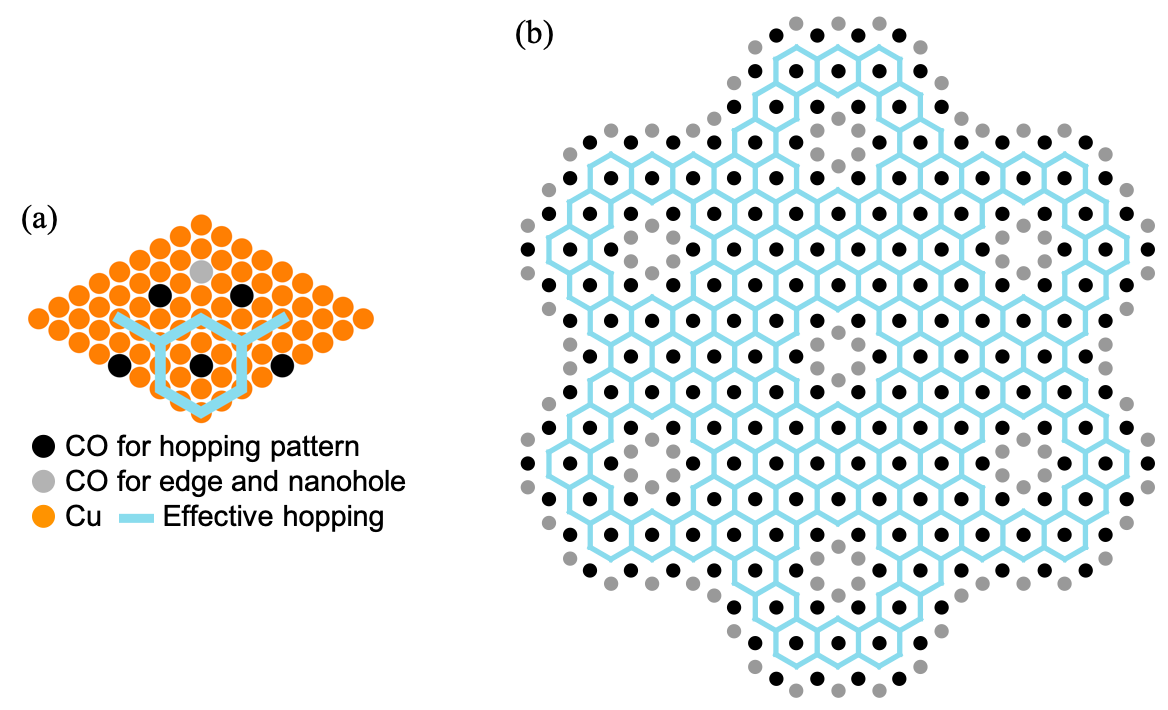}
		\caption{Schematic for molecular graphene.
		(a) CO molecules on Cu(111) surface. The CO molecules given by black dots form the effective hopping pattern while the CO molecules given by gray dots form the edge morphology and the nanoholes.
		(c) Molecular graphene for $3\sqrt{3}\times3\sqrt{3}$ triangular array of nanoholes, which consists of 306 CO molecules in total.
		}
		\label{fig:MG}
	\end{figure}

	As a proof-of-concept experiment, we suggest to realize the flake structure in terms of \textit{molecular graphene}, which is constructed by aligning CO molecules on Cu(111) surface with atomic precision by using the STM technique~\cite{Gomes2012}.
	Here we provide the design of molecular graphenes with $3\sqrt{3}\times3\sqrt{3}$ triangular array of nanoholes as shown in Fig.~\ref{fig:MG}.
	The positioning of CO molecules on the Cu(111) surface is displayed schematically in Fig.~\ref{fig:MG}(a), where besides
	the CO molecules used to form the hopping pattern (black dots), additional CO molecules are used to block undesired interactions with surrounding two-dimensional electron gas (gray dots)~\cite{freeney2020} to create the effective nanoholes and the edge morphology where all hexagonal unit cells remain intact.
	The total number of CO molecules used to construct the molecular graphene flake shown in Fig.~\ref{fig:MG}(b) consists of 306 CO molecules.
	Because our design structure is smaller than the molecular graphenes realized in a recent experiment where maximally 522 CO molecules were used~\cite{freeney2020}, the topological electronic states in the graphene flake proposed here is ready to be verified experimentally.

\section{Discussion and conclusion}

    In order to realize topological states in graphene with superstructures, nanohole arrays should be introduced in atomic precision.
    Advanced techniques for this purpose include nanoporous graphenes and graphene nanoribbons using molecular polymerization~\cite{Moreno2018,Song2025},
    the helium microscopy~\cite{Nakaharai2013} and graphene hydrogenization~\cite{Balog2010,Chen2018}. 
    Chemical synthesis based on precursors with heteroatoms~\cite{Kawai2018} is also promising.
    While in the present work we focus explicitly on graphene, the scheme conceived here is applicable to other Dirac materials, with the honeycomb patterned GaAs quantum wells~\cite{Wang2018} as a hopeful candidate.

To summarize, we investigate theoretically the nontrivial band topology for graphene with regular arrays of nanoholes with $C_{6v}$ symmetry, where an energy gap is opened at $\Gamma$ point of BZ due to intervalley interactions associated with the BZ folding from the pristine graphene.  Especially, for the case of $3\sqrt{3}\times3\sqrt{3}$ triangular array of nanoholes, which nanoholes are formed by removing six carbon atoms, we find a large energy band gap at $\Gamma$ point at the Fermi level and a band inversion between bands of opposite parities. In contrast to the simplest consideration based on the two bands close to the Fermi level, no topological edge states can be observed. In order to understand this behavior, we analyze the full VBs and find that, in addition to the above band inversion at the Fermi level, there are a bunch of VBs much below the Fermi level characterized by OAL which also accommodate an imbalance in parity indices. Therefore, the two top VBs can be considered as fragile topology carrying no topological edge states. It is revealed that the nontrivial band topology of the OAL manifests in two flat bands in the ribbon structure associated with localized electronic states at the ribbon edges. The present results exhibit rich topological behaviors in graphene derivatives waiting for experimental exploration and potentially useful for electronic device applications.

\addcontentsline{toc}{chapter}{Appendix A: Appendix section heading}
\section*{Appendix}

    Character tables and irreps for point groups at high-symmetric momenta in a $C_{6v}$-symmetric system are presented in Table~\ref{fig:table}.

 	\begin{table}[ht]
		\centering
		\caption{Irreps at high-symmetric momenta for a $C_{6v}$-symmetric system.
		Top panel: character tables for $C_{6v}$, $C_{3v}$, $C_{2v}$ and $C_s$ point groups, 
		with standard notations of irreps~\cite{Mulliken1933} in the first column and notations of irreps at the $\Gamma$, K, M and $\Sigma$~\cite{Elcoro2017} in the second column.
		Bottom panel: all possible irreps along the mirror-symmetric line $\Sigma$ in BZ at high-symmetric momenta for the two bands with degenerate states at the $\Gamma$ point in a $C_{6v}$-symmetric system~\cite{Sakoda2005}.
		}
        \includegraphics[width=0.95\textwidth]{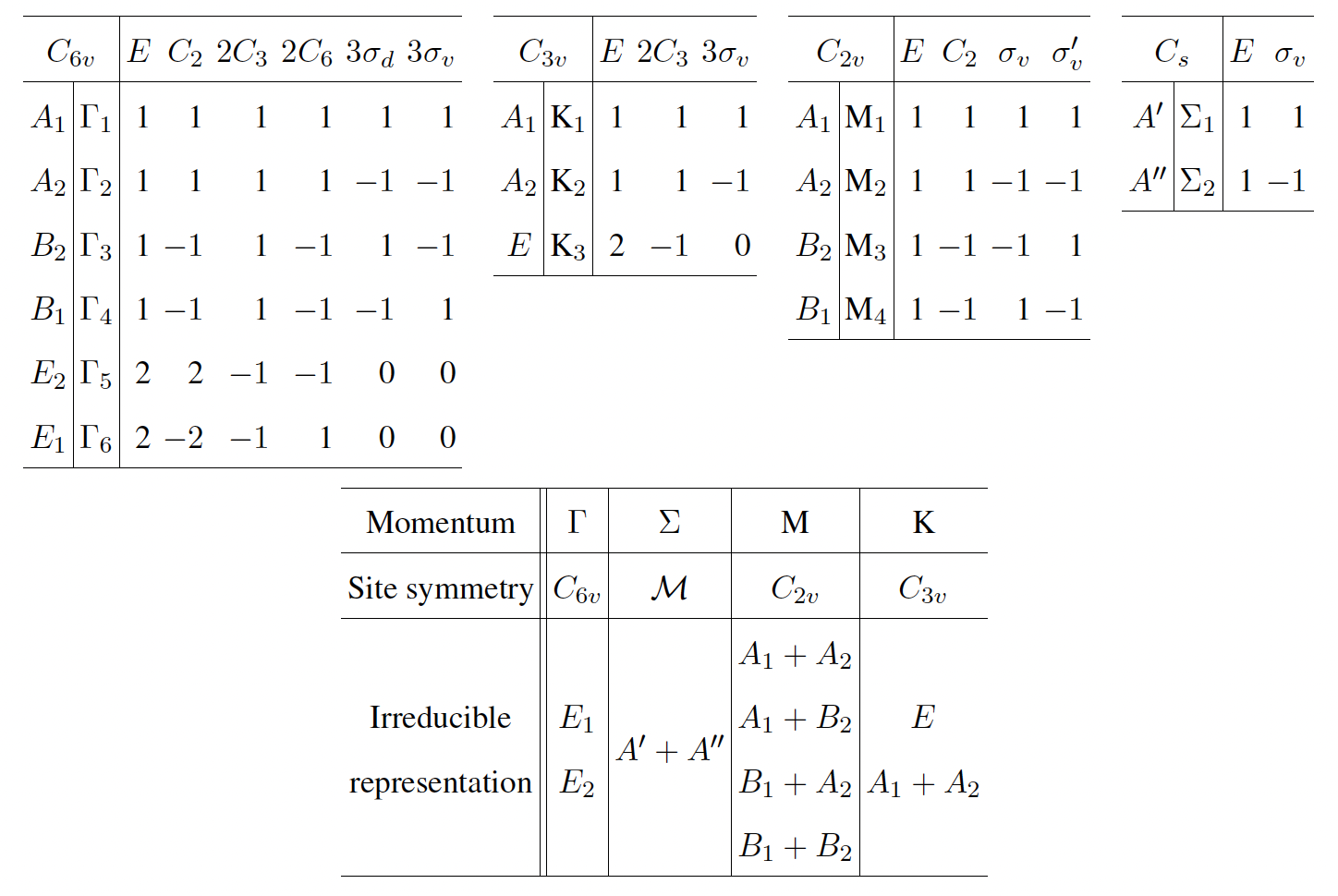}
		\label{fig:table}
	\end{table}

\section*{Data availability statement}
All data needed to evaluate the conclusions in the paper are present in the paper and/or the Supplementary Materials. Additional data related to this paper may be requested from the corresponding author.

\addcontentsline{toc}{chapter}{Acknowledgments}
\section*{Acknowledgments}
The authors are grateful to T. Kariyado for valuable discussions. This work is supported by Shanghai Science and Technology Innovation Action Plan (No. 24LZ1400800).

\bibliographystyle{iopart-num-modified} 
\bibliography{Topo_triangular}

%\end{CJK*}  %% end the Chinese environment
%\end{document}  %%% end document

\end{document}